\documentclass[12pt,epsf]{article}
\usepackage{amssymb,amsmath,latexsym,graphics, graphicx,epsfig}
\usepackage{latexsym}
\usepackage{amssymb}
\usepackage{epsfig,amsmath,graphics}
\usepackage{multirow}
\usepackage{appendix}
\usepackage{slashed}

\usepackage{cite}

\def\beq{\begin{equation}}
\def\eq{\end{equation}}
\def\eeq{\end{equation}}
\def\bea{\begin{eqnarray}}
\def\eea{\end{eqnarray}}

\def\centeron#1#2{{\setbox0=\hbox{#1}\setbox1=\hbox{#2}\ifdim
\wd1>\wd0\kern.5\wd1\kern-.5\wd0\fi
\copy0\kern-.5\wd0\kern-.5\wd1\copy1\ifdim\wd0>\wd1
\kern.5\wd0\kern-.5\wd1\fi}}
\def\ltap{\;\centeron{\raise.35ex\hbox{$<$}}{\lower.65ex\hbox{$\sim$}}\;}
\def\gtap{\;\centeron{\raise.35ex\hbox{$>$}}{\lower.65ex\hbox{$\sim$}}\;}

\def\chii0{\chi_i^0}
\def\chij0{\chi_j^0}

\def\beq{\begin{equation}}
\def\eq{\end{equation}}
\def\eeq{\end{equation}}
\def\barray{\begin{eqnarray}}
\def\earray{\end{eqnarray}}

\def\drawbox#1#2{\hrule height#2pt
        \hbox{\vrule width#2pt height#1pt \kern#1pt
               \vrule width#2pt}
               \hrule height#2pt}

\def\Fund#1#2{\vcenter{\vbox{\drawbox{#1}{#2}}}}
\def\Asym#1#2{\vcenter{\vbox{\drawbox{#1}{#2}
              \kern-#2pt       % line up boxes
              \drawbox{#1}{#2}}}}

\def\sym{\Fund{6.5}{0.4} \kern-.5pt \Fund{6.5}{0.4}}

\begin{document}

\begin{titlepage}

\begin{center}
\vspace*{-1cm}
 
\hfill RU-NHETC-2011-18 \\
\vskip 1.3in
{\LARGE \bf Recovering Particle Masses from  } \\
\vskip .1in \vspace{.1in} {\LARGE \bf  Missing Energy Signatures } \vskip
.1in \vspace{.1in} {\LARGE \bf with Displaced Tracks  }

\vskip 0.5in {\large Michael Park} and
{\large Yue Zhao}

\vskip 0.15in

{\em New High Energy Theory Center \\
Department of Physics \\
Rutgers University \\
Piscataway, NJ 08854}

\vskip 0.75in

\end{center}

\baselineskip=16pt

\begin{abstract}

\noindent

If physics beyond the Standard Model contains metastable new particle states, its signatures may manifest themselves at the LHC through the presence of displaced vertices or displaced tracks. Such signatures are common in low-scale supersymmetry breaking scenarios as well as a host of other well-studied new physics models. Here we analyze the kinematics of dual cascade decays with missing energy resulting from a stable non-interacting new particle state at the bottom of the decay chain. We find that if the final step of the decay chain involves a metastable particle, the resulting presence of displaced tracks provides strong constraints with which to solve for the unknown kinematic quantities lost through missing energy. In addition we develop techniques with which to recover all kinematic unknowns, even in situations where the unknown quantities outnumber the constraints. These techniques can all be performed using a very small number of events and can thus be applied to very early discovery level searches at the LHC.

\end{abstract}

\end{titlepage}

\baselineskip=17pt

\newpage

%
% --------------- Introduction -----------------------------
%

\section{Introduction}

The Large Hadron Collider (LHC), now in the early stages of discovery level searches, is currently probing the weak scale for signs of physics beyond the Standard Model (SM). Although a model independent approach to new physics searches should primarily involve searching for deviations from SM predictions of any kind, a specific discovery cannot be claimed without more detailed information about the processes that occur subsequent to the initial particle collisions. Therefore obtaining precise measurements of theoretical parameters, such as the mass spectrum of new particle states, is an endeavor of particular importance.

In this paper, we propose techniques for measuring particle masses from several different signatures containing missing transverse momentum. We assume that some heavy new particle states are pair produced and then participate in sequential two-body cascade decays that produce visible SM particles, until some effectively stable and non-interacting new particle is reached at the bottom of the decay chain. This is the canonical and well-studied ``dual cascade decay chain'' signature, well known for being the canonical signature of R-parity conserving supersymmetry (SUSY) models. None of the kinematic techniques discussed in this paper will rely on the fact that the cascade decay chains be supersymmetric in nature, thus all of these techniques may be applied generally to any BSM model that contains this topology as a signature. However, due to the familiarity with supersymmetric terminology, we will generically refer to the ``Lightest meta-Stable Particle'' as the ``LSP'' and the ``Next-to-Lightest meta-Stable Particle'' as the ``NLSP''. Our analysis will focus on a subset of these scenarios in which the last step of the cascade decay involves some long-lived new particle state that travels a finite distance before decaying in flight This will result in a signature of displaced vertices or displaced tracks in the detector.

The techniques to be described here are model independent which is fortunate since missing transverse momentum is a fairly generic feature of models for physics beyond the SM. This is because general phenomenological considerations often motivate new discrete symmetries, resulting in the presence of effectively non-interacting stable particle states. In the case of SUSY for example, R-parity is often invoked to exclude dangerous operators that can result in phenomenologically inconsistent effects like proton decay. In the case of extra dimensional models, it is the conservation of momentum along the extra dimension that will result in the pair production of Kaluza-Klein (KK) states and subsequently guarantee the stability of the lightest KK mode. One can even invoke cosmological arguments like the ``WIMP Miracle'' calculations to argue that missing energy signatures might be a generic phenomenologically desirable feature of models for new physics at the weak scale. The presence of metastable new particle states is also fairly common and can arise in supersymmetric models with low scale SUSY breaking or scenarios where R-parity conservation is only approximate. In this paper, we address the question of whether or not it is possible, under any circumstances, to recover all of the kinematic information lost through missing energy on an event-by-event basis. 

If all of the final state particles from a given collision are visible through the detector, then the measurement of on-shell particle masses can easily be performed through the straightforward reconstruction of a mass resonance peak. However, if one or more of the final state particles are effectively stable and non-interacting, then the situation is much more challenging. In particular, particle masses cannot be calculated directly on a mass peak resonance since crucial kinematic quantities cannot be measured. In response to this issue, many general techniques have been developed for performing indirect measurements of particle masses through cleverly constructed kinematic variables \cite{Hinchliffe:1996iu,Lester:1999tx,Barr:2003rg,Tovey:2008ui,Matchev:2009ad,Konar:2010ma}. In particular, the author in \cite{Kim:2009si} introduces a very generic method for constructing such variables via phase space singularity structures. Many studies have also been performed based on kinematics specific to the canonical cascade decay chain \cite{Bachacou:1999zb,Jackson:2001iy,Covi:2001nw,Allanach:2000kt,Gjelsten:2004ki, Kawagoe:2004rz,Lester:2005je,ArkaniHamed:2005px,Baumgart:2006pa,Butterworth:2007ke, Cheng:2007xv,Nojiri:2007pq,Horsky:2008yi,Barr:2009wu,Cohen:2010wv,NNOMET}. In general, novel kinematic structures that characterize an event can often be used to reconstruct lost information. For example, \cite{Kawagoe:2003jv,Meade:2010ji} also discuss a long-lived NLSP, and the use of timing information to perform such reconstructions. For different topologies of the decay chain, \cite{Barr:2010zj} provides a comprehensive review. The drawback to most of these methods is the fact that most of the kinematic variables that can be constructed to provide an indirect mass measurement, require a very large number of events for telling features to become practically visible in statistical distributions. These methods would therefore be difficult to utilize during early discovery level searches. 

First, using our assumptions we will show that one can write down an expression for the 3-momenta of each LSP as a function of the direction of the 3-momenta of each NLSP. This unit vector can then be written in terms of the locations of the secondary vertices. We will then follow with a description of some novel methods for reconstructing particle masses using this information. Examples of explicit mass reconstruction will be performed using Monte Carlo parton-level data, highlighting the effectiveness of these methods in some of the diverse topologies that can occur within the cascade. Finally we will conclude with a discussion of how this relates to current SUSY searches being performed at the LHC. We argue then, that the optimal strategy for ``searching under the lamp post'' during these early runs will be to search for signatures with two displaced vertices (or two displaced tracks in situations where the exact location of the secondary vertices cannot be measured).

\section{Counting the Unknowns}

\begin{figure}[!ht]
\begin{center}
\includegraphics[width=100mm]{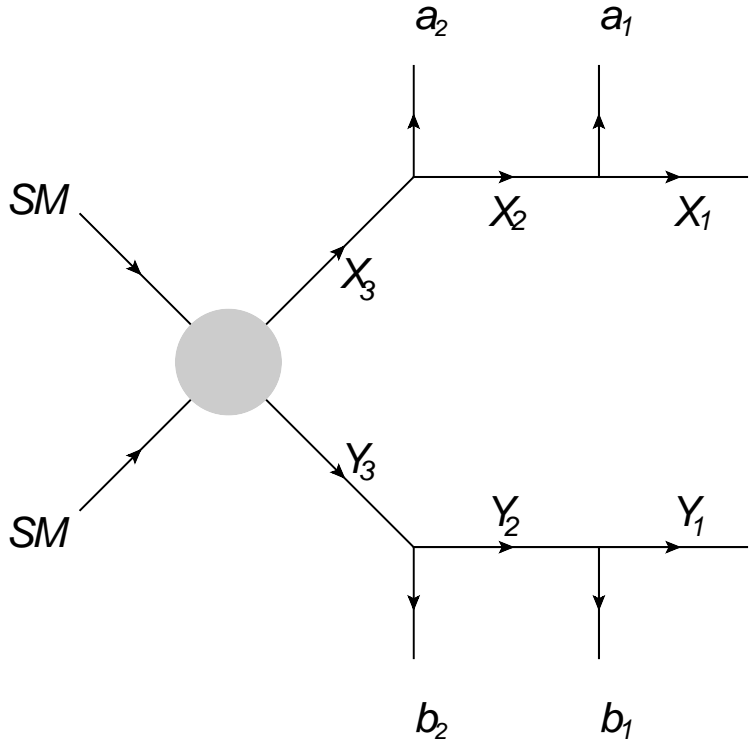}
\caption{The decay topology for the canonical dual cascade decay chain. Two heavy new particle states ($X_3$ and $Y_3$) are pair produced. They then cascade down to a pair of non-interacting stable particles ($X_1$ and $Y_1$) shooting off visible SM particles ($a_1$, $b_1$, $a_2$ and $b_2$) along the way.}
\label{proc}
\end{center}
\end{figure}

Let us denote the stable LSP particles by $X_1$ and $Y_1$, their mother NLSP particles as $X_2$ and $Y_2$ and the final visible SM particles as $a_1$ and $b_1$. Figure [\ref{proc}] shows a diagram of a typical event. Since we assume that the 4-momenta of the two LSP's are not measurable, each event yields $8$ unknown quantities. The transverse missing momentum is given by the vector sum of the LSP 3-momenta projected onto the transverse plane. Since this plane is 2-dimensional, a missing transverse momentum measurement eliminates $2$ degrees of freedom bringing the number of unknowns down to $6$. In order to construct constraint equations with which to solve for these unknowns, we follow the work of \cite{Cheng:2008mg,Cheng:2009fw,Webber:2009vm} and assume some symmetry between the two sides of the decay chains. For example, if we assume that $m_{X_2} = m_{Y_2}$ then we can use the fact that $(p_{X_1}^\mu + p_{a_1}^\mu)^2 = (p_{Y_1}^\mu + p_{b_1}^\mu)^2$ as a constraint with which to eliminate one of the unknown momentum components. 

Let $k$ denote the number of such equations we can construct. Since all of the unknown quantities involve components of the LSP 4-momenta, $k$ can be viewed as the number of masses starting from the bottom of one decay chain but excluding the LSP, that we assume to be equal to the masses on the opposite side of the decay chain. Utilizing these constraints, the number of unknowns can be reduced to $6 - k$. It is important to keep in mind however, that such an exact relationship between the masses of these particles only holds in the very narrow width limit. In general, the true kinematically reconstructed masses will lie on the distribution of some mass peak resonance and the equality of the masses will only be approximately true. This will affect both the accuracy of the mass measurement as well as potentially the existence of solutions to the constraint equations. We will return to a more detailed discussion of this in the body of the paper.

If we assume that we have access to $m$ events with the same topology then we can use the equality of masses across events to further constrain the problem as done in \cite{Cheng:2008mg,Cheng:2009fw,Webber:2009vm}. For the first event we counted $6 - k$ unknowns. Each additional event contributes another $6 - k$ unknowns but if we enforce the equality of masses across different events then we should subtract off another factor of $k$. Thus each additional event contributes $6 - 2k$. For $m$ events, the total number of unknowns is $6 - k + (m - 1)(6 - 2k) = 6m - 2km + k$. The condition which must be satisfied in order to properly constrain the problem is thus clearly $6m - 2km + k \leq 0$.

Given our assumptions that the NLSP is the only particle in the spectrum with a finite and measurable decay length, an accurate measurement of the locations of the displaced vertices can be used to provide additional constraints. Here we assume that all of the decays occur on a microscopic length scale before the NLSP's travel a finite macroscopic distance and decay to a pair of invisible LSP's and a pair of visible SM particles. This implies that the direction of the NLSP 3-momentum is equal to the unit vector pointing in the direction of the secondary vertex. The NLSP unit 3-momentum contains two degrees of freedom, thus an accurate measurement of two displaced vertices will allow us to subtract off another $4m$ unknowns. In some situations, it will not be possible to measure the locations of the displaced vertices and only the trajectories of the displaced tracks will be visible. In these situations, the locations of the secondary vertices can be constrained to lie on the trajectories of the displaced tracks and can be parameterized by one number thus removing $2m$ unknown quantities.

\section{Parameterizing the Unknowns}

In this section we propose a parameterization of the unknown quantities that makes the utility of displaced vertices and displaced tracks maximally transparent. More specifically, we will show that it will be possible to write down an expression for the 3-momenta of each LSP that depends only the location of the displaced vertices. Throughout this analysis let us assume that the 4-momenta of the visible Standard Model particles $a_1$ and $b_1$ can be measured accurately. Let us restrict our attention to one side of the decay chain and denote the 4-momenta for particle $X_1$, $X_2$ and $a_1$ as in Eq. [\ref{fourvecparam}]

\beq
p_{X_1} = \begin{pmatrix}
E_{X_1} \\
|\vec{p}_{X_1}| \hat{p}_{X_1} \\ \end{pmatrix}  \quad \text{;} \quad 
p_{X_2} = \begin{pmatrix} E_{X_2} \\
|\vec{p}_{X_2}| \hat{p}_{X_2} \\ \end{pmatrix} \quad \text{;} \quad
p_{a_1} = \begin{pmatrix}
E_{a_1} \\
|\vec{p}_{a_1}| \hat{p}_{a_1} \\ \end{pmatrix}
\label{fourvecparam}
\eq

\begin{figure}[!ht]
\begin{center}
\includegraphics[width=70mm]{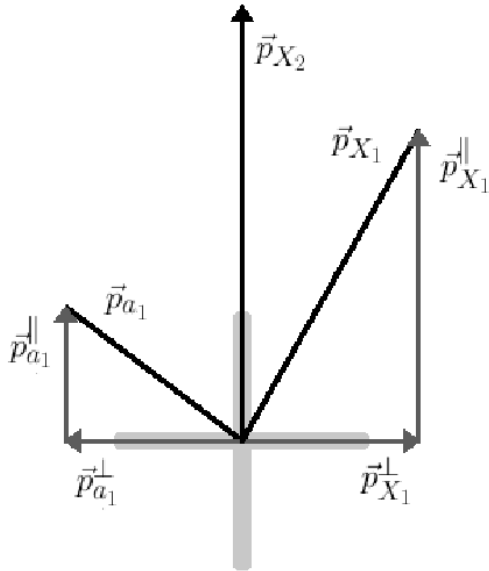}
\caption{A decomposition of the final decay products into their components parallel and orthogonal to particle $X_2$. Note that the component of $\vec{p}_{X_1}$ orthogonal to the direction of the NLSP, is equal in magnitude and opposite in direction to the component of $\vec{p}_{a_1}$ orthogonal to the direction of the NLSP.}
\label{decomp}
\end{center}
\end{figure}

To isolate the unknown quantities, it is useful to decompose the 3-momenta of particles $a_1$ and $X_1$ in terms of their components parallel and orthogonal to the momentum of particle $X_2$ as in Figure [\ref{decomp}]. For notational convenience, let us define the projection symbol as in Eq. [\ref{projsym}]

\beq
\mathbb{P}^i_j \equiv \vec{p}_i \cdot \hat{p}_j
\label{projsym}
\eq

\noindent This denotes the projection of the 3-momentum of particle $i$ along the direction of the 3-momentum of a different particle $j$. In this basis and with this notation we can decompose $\vec{p}_{a_1}$ into its components parallel $\vec{p}^\parallel_{a_1} = \mathbb{P}^{a_1}_{X_2} \hat{p}_{X_2}$ and orthogonal $\vec{p}^\perp_{a_1} = \vec{p}_{a_1} - \mathbb{P}^{a_1}_{X_2} \hat{p}_{X_2}$ to particle $X_2$. Conservation of momentum then allows us to immediately write down the orthogonal component of the 3-momentum of particle $X_1$ as $\vec{p}^\perp_{X_1} = -\vec{p}^\perp_{a_1} = \mathbb{P}^{a_1}_{X_2} \hat{p}_{X_2} - \vec{p}_{a_1}$. The magnitude of the component of the 3-momentum of particle $X_1$ along the direction of $X_2$ remains unknown. In this paper we will denote the magnitude of this unknown as $c_1 \equiv \mathbb{P}^{X_1}_{X_2}$ so the component of the LSP parallel to the direction of the NLSP can be expressed as $\vec{p}^\parallel_{X_1} = c_1 \hat{p}_{X_2}$. Since the other side of the decay chain is subject to identical kinematic considerations, the 3-momentum of each LSP is given by Eq. [\ref{threevecexp}]

\beq
\vec{p}_{X_1} = (\mathbb{P}^{a_1}_{X_2} + c_1)\hat{p}_{X_2} - \vec{p}_{a_1} \quad \text{and} \quad \vec{p}_{Y_1} = (\mathbb{P}^{b_1}_{Y_2} + c_2)\hat{p}_{Y_2} - \vec{p}_{b_1}
\label{threevecexp}
\eq

Let $\alpha = 1,2$ be indices parameterizing a basis in the two-dimensional transverse plane. The experimentally measured missing transverse momentum  $\vec{\slashed{p}}^T_\alpha$ contains two degrees of freedom and is restricted to the transverse plane. Since by assumption, the missing transverse momentum in this scenario is taken from the vector sum of the 3-momenta of the two LSP's, it can be calculated as the sum of contributions from each LSP as in Eq. [\ref{metcond}]

\beq
\vec{\slashed{p}}^T_\alpha = \vec{p}^{X_1}_\alpha + \vec{p}^{Y_1}_\alpha = (\mathbb{P}^{a_1}_{X_2} + c_1) \hat{p}^{X_2}_\alpha + (\mathbb{P}^{b_1}_{Y_2} + c_2) \hat{p}^{Y_2}_\alpha - \vec{p}^{a_1}_\alpha - \vec{p}^{b_1}_\alpha
\label{metcond}
\eq

\noindent These two equations can then be used to solve for $c_1$ and $c_2$ as in Eq. [\ref{c2}]

\[
c_1 = \frac{(p^{a_1}_\alpha p^{Y_2}_\beta + p^{b_1}_\alpha p^{Y_2}_\beta + \slashed{p}_\alpha p^{Y_2}_\beta)\epsilon^{\alpha \beta}}{p^{X_2}_\alpha p^{Y_2}_\beta \epsilon^{\alpha \beta}} - \mathbb{P}^{a_1}_{X_2}
\]
\beq
c_2 = \frac{(p^{a_1}_\alpha p^{X_2}_\beta + p^{b_1}_\alpha p^{X_2}_\beta + \slashed{p}_\alpha p^{X_2}_\beta)\epsilon^{\alpha \beta}}{p^{Y_2}_\alpha p^{X_2}_\beta \epsilon^{\alpha \beta}} - \mathbb{P}^{b_1}_{Y_2}
\label{c2}
\eq

\noindent Here $\epsilon^{\alpha \beta}$ is the totally antisymmetric $2 \times 2$ tensor. The key result here is that an accurate measurement of the missing transverse momentum will allow us to write down the 3-momentum of each LSP as a function of the direction of the NLSP 3-momenta by plugging Eq. [\ref{c2}] into Eq. [\ref{threevecexp}]. The result is summarized by Eq. [\ref{dirparam}]

\[
\vec{p}_{X_1} \rightarrow \vec{p}_{X_1} (\hat{p}_{X_2}, \hat{p}_{Y_2})
\]
\beq
\vec{p}_{Y_1} \rightarrow \vec{p}_{Y_1} (\hat{p}_{X_2}, \hat{p}_{Y_2})
\label{dirparam}
\eq

Let us denote the location of the two secondary vertices by 3-vectors in the Cartesian coordinates of the lab frame $\vec{r}_X$ and $\vec{r}_Y$. Here the subscripts $X$ and $Y$ correspond to the location of the decays of particles $X_2$ and $Y_2$. Note that given our assumptions $|\vec{r}_X| = d_X$ is simply the distance traveled by particle $X_2$ before decaying while $|\vec{r}_Y| = d_Y$ is the distance traveled by particle $Y_2$ before decaying, assuming all other decays are prompt. Now recall our initial assumption that the decay length of particles $X_2$ and $Y_2$ are the only decay lengths that are measurably large. Then subsequent to the initial collision, a cascade will occur on some microscopic length scale before the NLSP's travel a finite macroscopic distance and decay to a pair of invisible LSP's and a pair of visible SM particles. This implies that the direction of the NLSP 3-momentum is equal to the unit vector pointing in the direction of the secondary vertex. The exact relationship is $\hat{p}_{X_2} = \vec{r}_X / |\vec{r}_X|$. Therefore in actuality we have derived an expression for the LSP 3-momenta that depends only on the location of the secondary vertices as in Eq. [\ref{vertparam}]

\[
\vec{p}_{X_1} \rightarrow \vec{p}_{X_1} (\vec{r}_{X}, \vec{r}_{Y})
\]
\beq
\vec{p}_{Y_1} \rightarrow \vec{p}_{Y_1} (\vec{r}_{X}, \vec{r}_{Y})
\label{vertparam}
\eq

In some situations, the displaced vertices may not be directly measurable and only the trajectories of the displaced tracks may be extracted. However, it may be inferred that the displaced vertices must lie somewhere along the path of the displaced tracks. We may thus parameterize the location of the displaced vertices according to their location along the beam axis. Let $z_X$ and $z_Y$ denote the location along the z-axis of $\vec{r}_X$ and $\vec{r}_Y$ respectively and let us set the location of the primary vertex to be $z=0$. Indeed if we denote the location of particle $a_1$'s collision with the tracker by $\vec{r}_0 = (x_0, y_0, z_0)$, then an exact functional form for $\vec{r}_X(z_X)$ is given by Eq. [\ref{trackparam}]

\beq
\vec{r}_X (z_X) = \begin{pmatrix}
x_0 + (p^{a_2}_{\hat{x}} / p^{a_1}_{\hat{z}}) (z_X - z_0) \\
y_0 + (p^{a_2}_{\hat{y}} / p^{a_1}_{\hat{z}}) (z_X - z_0) \\
z_X \\ \end{pmatrix}
\label{trackparam}
\eq

\noindent This will allow us to derive an expression for the LSP 3-momenta that depends only on the location of the secondary vertices along the beam axis as in Eq. [\ref{zparam}]

\[
\vec{p}_{X_1} \rightarrow \vec{p}_{X_1} (z_X, z_Y)
\]
\beq
\vec{p}_{Y_1} \rightarrow \vec{p}_{Y_1} (z_X, z_Y)
\label{zparam}
\eq

\noindent From this parameterization we can explicitly see the dependence of the 3-momentum of each missing particle on the locations of the displaced vertices or the trajectories of the displaced tracks. Now that it is clear how such measurements can be used to reduce the number of unknowns and further constrain the kinematics of this decay topology, we move on to some practical examples.

\section{Examples with a Massless LSP}

From the counting arguments given in the introduction, we found that for $m$ events and $k$ constraint equations, the total number of unknown quantities was equal to $6m - 2km + k$. In principle, the problem is simply a matter of solving for enough constraint equations to obtain a unique solution for all unknown quantities. In practice however, the contraint equations are highly nonlinear and generically contain multiple solutions. As a result, a confident mass measurement should really involve the analysis of a number of events greater than the minimum required to properly constrain the problem. We will now explore a few specific examples. For concreteness, we will start with an analysis of selected benchmark points for multi-lepton searches inspired by scenarios with general gauge mediated SUSY breaking (GMSB). In such scenarios, where the scale of SUSY breaking is sufficiently low, the LSP is an effectively massless gravitino. With the mass of the LSP set to zero, $2m$ unknowns are removed from the problem resulting in a total number given by Eq. [\ref{masslesscount}]

\beq
\text{Number of Unknowns for Massless LSP Scenario = } 4m - 2km + k
\label{masslesscount}
\eq

\subsection{Measurable Displaced Vertices}

\begin{figure}[!ht]
\begin{center}
\includegraphics[width=100mm]{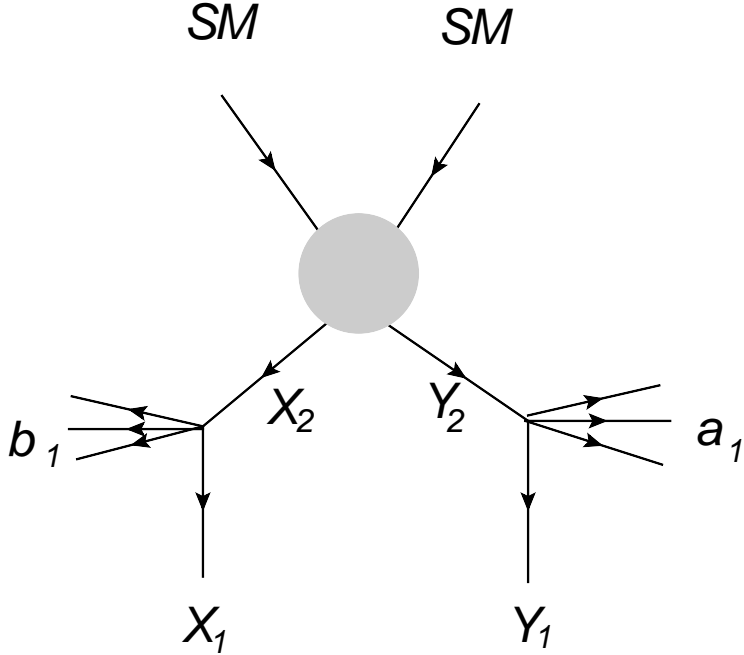}
\caption{A dual cascade chain with prompt Standard Model particle decays. If the visible SM particle decays promptly and if all of its decay products are visible, then the trajectories of its decay products may be traced back to the displaced vertex.}
\label{caseone1}
\end{center}
\end{figure}

If $a_1$ and $b_1$ each decay promptly to two or more visible particles, it will be possible to experimentally trace back the track of each decay product and reconstruct the position of the secondary vertex. The momenta $p_{a_1}$ and $p_{b_1}$ can then be computed through the sum of 4-momenta of their respective decay products, assuming none of them contribute to the missing transverse momentum. The situation is depicted in Figure [\ref{caseone1}]. In this case we can directly measure the quantities $\hat{p}_{X_2} = \vec{r}_X / |\vec{r}_X|$ and $\hat{p}_{Y_2} = \vec{r}_Y / |\vec{r}_Y|$ and thus completely solve for the 3-momenta of particles $X_1$ and $Y_1$. Since these particles are massless by assumption, a measurement of the 3-momenta is equivalent to a measurement of the full 4-momenta. Therefore in situations where the LSP is massless, a simple measurement of the locations of the displaced vertices already recovers all of the information lost through missing energy. In terms of the unknowns we see that substituting one event $m=1$ into Eq. [\ref{masslesscount}] gives us $4-k$. A measurement of the displaced vertices removes exactly $4$ unknowns, which means that the condition for total kinematic recovery is already met for $k=0$. This example is thus trivial and will not be discussed further. 

\subsection{Measurable Displaced Tracks}

If $a_1$ and $b_1$ are stable and hit the detector, then $p_{a_1}$ and $p_{b_1}$ can be directly measured. In this scenario, the exact location of the secondary vertices cannot be measured but one can constrain their location to points along the displaced tracks of particles $a_1$ and $b_1$. Parameterizing the 3-momentum $\vec{p}_{X_1}$ by $z_X$ and $\vec{p}_{Y_1}$ by $z_Y$ removes $2m$ unknown quantities from Eq. [\ref{masslesscount}] bringing the requirement for total kinematic recovery down to $2m - 2km + k \leq 0$. Acheiving this with $m=1$ event requires that $k \geq 2$ so we will use the fact that $m_{X_3} = m_{Y_3}$ and $m_{X_2} = m_{Y_2}$ in order to measure the particle masses. Our canonical example for this scenario, depicted in Figure [\ref{casetwo1}], comes from GMSB. Here we consider the case where two partons collide resulting in the pair production of two right-handed squarks. Each squark decays to Bino-like neutralinos $X_3$ and $Y_3$, emitting jets in the process. Each neutralino then decays to right-handed sleptons $X_2$ and $Y_2$, emitting leptons $a_2$ and $b_2$ in the process. Finally the right-handed sleptons decay to the LSP gravitinos $X_1$ and $Y_1$, emitting additional leptons $a_1$ and $b_1$ in the process. The relevant part of the spectrum is summarized in the following table:

\begin{figure}[!ht]
\begin{center}
\includegraphics[width=100mm]{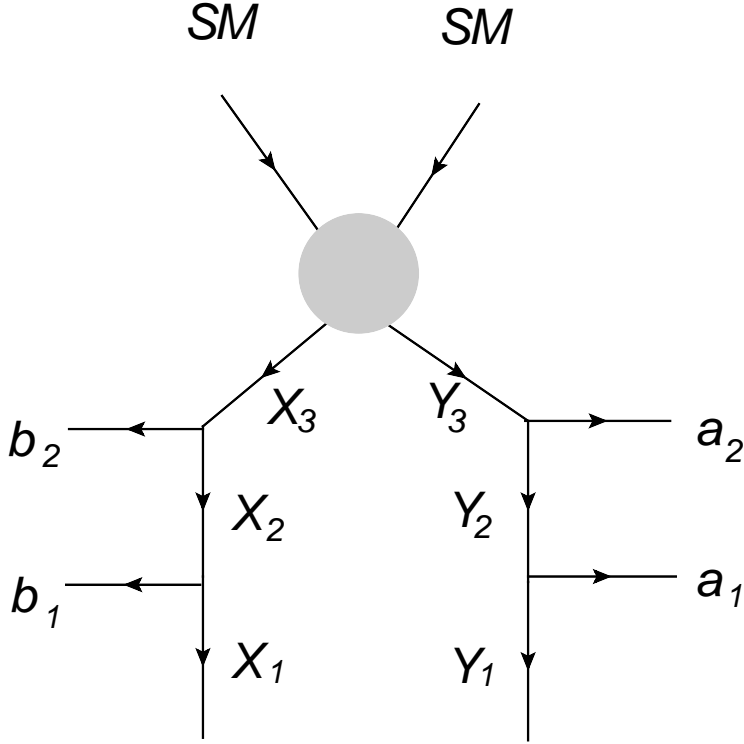}
\caption{A dual cascade chain with two  SM ``legs''. This is the scenario where $k=3$ so the decay chain must have at least two SM legs in order for this technique to be effective.}
\label{casetwo1}
\end{center}
\end{figure}

\begin{center}
\begin{tabular}{|c|c|c|}
\hline
Particle & Symbol & Mass \\
\hline
Bino & $\tilde{B}$ & $199$ GeV \\
Right-handed Slepton & $\tilde{l}_R$ & $107$ GeV \\
Gravitino & $\tilde{G}$ & $0$ GeV \\
\hline
\end{tabular}
\end{center}

Another point to keep in mind is that the further up we go in the decay chain, the higher the chance for combinatoric confusion among the visible particles, labeled in this example by $a_1$, $a_2$, $b_1$ and $b_2$. In general it may not always be possible to identify the correct particle with its correct position within a given decay chain. If this is the case then all possibilities should be considered which will result in a larger multiplicty of solutions. A slightly larger data sample may then be required in order to make a definitive mass measurement by finding a common value for the masses. Since the visible SM particles are leptons, we will treat them as effectively massless. The relevant formulae are then given in Eq. [\ref{masses1}] with the expressions for $c_1$ and $c_2$ given by Eq. [\ref{c2}].

\[
m_{X_3}^2 = 2 (E_{a_1} + E_{a_2}) \sqrt{c_1^2 - (\mathbb{P}^{a_1}_{X_2})^2 + \vec{p}_{a_1}^2} - 2(c_1 + \mathbb{P}^{a_1}_{X_2}) (\mathbb{P}^{a_1}_{X_2} + \mathbb{P}^{a_2}_{X_2}) + 2 \vec{p}_{a_1}^2 + 2 E_{a_1} E_{a_2}
\]
\[ 
m_{Y_3}^2 = 2 (E_{b_1} + E_{b_2}) \sqrt{c_2^2 - (\mathbb{P}^{b_1}_{Y_2})^2 + \vec{p}_{b_1}^2} - 2(c_2 + \mathbb{P}^{b_1}_{Y_2}) (\mathbb{P}^{b_1}_{Y_2} + \mathbb{P}^{b_2}_{Y_2}) + 2 \vec{p}_{b_1}^2 + 2 E_{b_1} E_{b_2} 
\]
\[
m_{X_2}^2 =2 E_{a_1} \sqrt{c_1^2 + \vec{p}_{a_1}^2 - (\mathbb{P}^{a_1}_{X_2})^2} - 2(\mathbb{P}^{a_1}_{X_2} + c_1)\mathbb{P}^{a_1}_{X_2} + 2\vec{p}_{a_1}^2
\]
\beq
m_{Y_2}^2 = 2 E_{b_1} \sqrt{c_2^2 + \vec{p}_{b_1}^2 - (\mathbb{P}^{b_1}_{Y_2})^2} - 2(\mathbb{P}^{b_1}_{Y_2} + c_2)\mathbb{P}^{b_1}_{Y_2} + 2\vec{p}_{b_1}^2
\label{masses1}
\eq

\noindent In practice, we are using two equations to solve for two unknowns $m_{X_3}(z_X, z_Y) = m_{Y_3}(z_X, z_Y)$ and $m_{X_2}(z_X, z_Y) = m_{Y_2}(z_X, z_Y)$. The calculation of unknown particle masses $m_{X_2}$ and $m_{X_3}$ in this scenario is presented here in the table with incorrect and correct solutions separated by columns:

\begin{center}
\begin{tabular}{|c|c|c|}
\hline
Event & correct (Bino, Slepton) &  wrong (Bino, Slepton)  \\
\hline
1 & (202, 108)& (467, 290) \\
2 & (199, 107)& (191, 93) \\
3 & (205, 111)& Null \\
4 & (200, 109)& (405, 346) \\
5 & (200, 108)& (209, 123),(490, 254) \\
\hline
\end{tabular}
\end{center}

\noindent Using $\mathcal{O}(few)$ events we see that the correct solutions can be separated from the incorrect solutions by their sheer multiplicity. %The following is a plot of parameter space with the correct answers denoted by a circle and erroneous solutions denoted by an $\times$:

\subsection{Can We Do Better?}

\begin{figure}[!ht]
\begin{center}
\includegraphics[width=100mm]{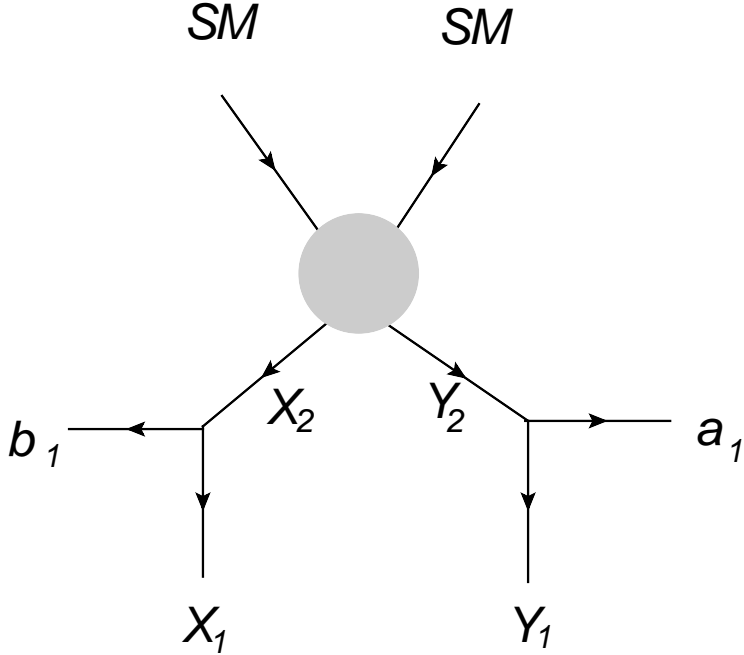}
\caption{A dual cascade chain with one Standard Model ``leg''. This is the most general possible scenario with the least amount of specificity. We only require that two equal mass NLSP's exist in the decay chain.}
\label{casethree1}
\end{center}
\end{figure}

It may be argued that the equality of masses following from the condition $k=2$ is too specific. Indeed if there was a way to measure the particle mass spectrum without demanding $m_{X_3} = m_{Y_3}$, these techniques would gain a lot in generality and become useful in a far wider range of possible new physics scenarios. Thus a natural next step would be to see if it would be possible, under any circumstances, to measure particle masses under the condition $k=1$ as depicted in Figure [\ref{casethree1}]. As demonstrated in the previous section, in scenarios with a massless LSP where the trajectories of the displaced tracks are known, the requirement for total kinematic recovery is  $2m - 2km + k \leq 0$. Solving for $k$ in terms of $m$ gives the expression $k \geq 2m/(2m-1)$. Clearly as $m \rightarrow \infty$,  $k \rightarrow 1$ asymptotically but the condition $k=1$ cannot be satisfied for any value of $m$. Naively this implies that it would not be possible to measure the particle masses given this assumption. Here we present a technique that defies this apparent restriction and demonstrate a particle mass measurement technique using only the condition $k=1$.

For situations in which the number of unknown quantities is larger than the number of constraint equations available, there exists a novel and unorthodox method of extracting particle masses using a relatively small number of events. The idea behind this method utilizes the fact that even in situations where the number of constraints is not large enough to specify a unique solution to all of the unknown quantities, it may be large enough to reduce the space of solutions down to a lower dimensional subspace where the solution may be inferred. Our toy model is taken again from a GMSB scenario. The process under consideration starts with the direct pair production of right-handed sleptons labeled here by $X_2$ and $Y_2$. The sleptons then decay to LSP gravitinos $X_1$ and $Y_1$, emitting a leptons $a_1$ and $b_1$ in the process. The relevant part of the mass spectrum is summarized in the following table:

\begin{center}
\begin{tabular}{|c|c|c|}
\hline
Particle & Symbol & Mass \\
\hline
Right-handed Slepton & $\tilde{l}_R$ & $107$ GeV \\
Gravitino & $\tilde{G}$ & $0$ GeV \\
\hline
\end{tabular}
\end{center}

The central challenge associated with this example is that there are two unknown quantities $z_X$ and $z_Y$ but only one mass constraint equation $m_{X_2}(z_X,z_Y) = m_{Y_2}(z_X,z_Y)$, which means that a unique solution cannot be obtained. However, this constraint allows us to express $z_X$ as a function of $z_Y$, which we may then use to write down an expression for the mass of a particle in terms of one variable $m_{X_2} (z_X)$. With this one-to-one map from $z_X$ to $m_{X_2}$, the space of possible solutions has been reduced to a one-dimensional subspace (i.e. a line) and the true value of $m_{X_2}$ must exist as an element of this subspace. 

Recall that all of the unknown quantities could be parameterized by the direction of the NLSP's $\hat{p}_{X_2}$ and $\hat{p}_{Y_2}$. Recall further that the direction of an NLSP is given by the location of its secondary vertex $\hat{p}_{X_2} = \vec{r}_X / |\vec{r}_X|$, which is restricted to lie somewhere along the trajectory of the associated displaced track. Recall finally, that a secondary vertex can thus be parameterized by its location along the beam axis $\vec{r}_X \rightarrow \vec{r}_X (z_X)$. The powerful observation here is the fact that as the hypothesized location of the displaced vertex along the beam axis approaches infinity ($z_X \rightarrow \infty$), the direction of the NLSP will asymptotically approach some fixed unit vector ($\hat{p}_{X_2} \rightarrow \hat{p}_{const}$). This means that as $z_X \rightarrow \infty$, the corresponding value of $m_{X_2} (z_X)$ will asymptotically approach some fixed number. In other words for the function $m_{X_2} (z_X)$, the domain $z_X \in (-\infty, \infty)$ maps to a closed finite range for $m_{X_2}$, and the correct value of $m_{X_2}$ will always be contained in this range. If we plot the elements of this range in a histogram over a small number of events, the histogram will peak around the correct solution since it is an element of every set and should thus have the highest multiplicity across events.

In principle, the correct values for $z_X$ and $z_Y$ can take on any arbitrary value. Since the decay distance of particles has the form of an exponentially decaying function, hypotheses for the location of the displaced vertex that are closer to the primary vertex should carry more weight than ones that are farther away. In order to attenuate contributions from unlikely vertex locations and increase the efficiency of our analysis, we scan the trajectory of the displaced track and assign a weight to each point accordingly. The weighting function is given by Eq. [\ref{weightfunc}]
 
\beq f[l]=\frac{e^{-l/l_0}}{g(d_{\perp})} \label{weightfunc} \eq

\begin{figure}[!ht]
\begin{center}
\includegraphics[width=100mm]{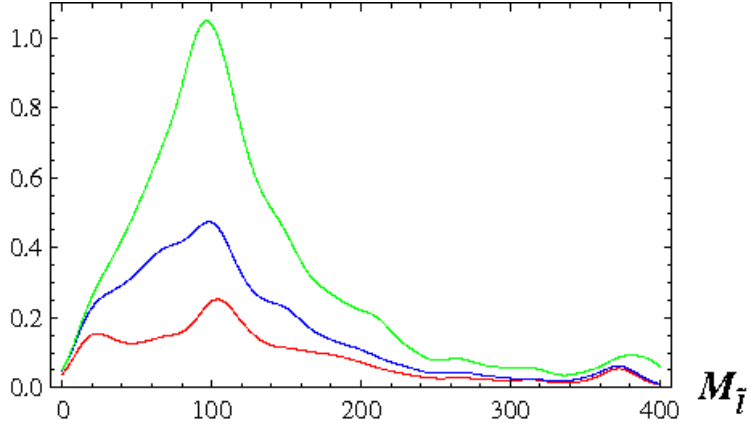}
\caption{Results of the likelihood fit. The red curve indicates an example with 15 events. The blue curve indicates an example with 30 events. The green curve indicates and example with 60 events}
\label{trickfit}
\end{center}
\end{figure}

Here $l$ is the distance between the point on the displaced track and the primary vertex and $l_0$ is the characteristic decay length of the NLSP. A more detailed discussion of this can be found in Appendix A. Note that we need as input only the rough order of this decay length which can be derived by looking at the distribution of displaced tracks as described in the Appendix B. The result of this weighted histogram is shown in Figure [\ref{trickfit}]. As we can see, this histogram quickly peaks at the value of the correct slepton mass of $107$ GeV.

\section{Examples with a Massive LSP}

Recall again from the introduction, that for $m$ events and $k$ constraint equations, the general scenario with a massive LSP resulted in a counting of unknown quantities given by Eq. [\ref{massivecount}]. In this section we will study such examples, that typically arise in the supersymmetric context when SUSY is broken at the Planck scale via gravity-mediation. The techniques described in this section will all be a straightforward demonstration of matching constraint equations with unknowns.

\beq
\text{Number of Unknowns for Massive LSP Scenario = } 6m - 2km + k
\label{massivecount}
\eq 

\subsection{Measurable Displaced Vertices}

Just as it was with the massless LSP, the requirement for measurable displaced vertices is that the final visible SM particles $a_1$ and $b_1$ must each decay promptly to two or more visible particles as depicted in Figure [\ref{caseone1}]. The measurement of displaced vertices will again provide us with a complete measurement of the LSP 3-momenta $\vec{p}_{X_2}$ and $\vec{p}_{Y_2}$. The difference is that now the mass of the LSP remains an unknown quantity in the LSP 4-momenta. 

In terms of our counting exercise, the measurement of displaced vertices subtracts $4m$ unknown quantities from Eq. [\ref{massivecount}] bringing the total number of unknowns down to $2m - 2km + k$. If we are interested in solving for all masses on an event-by-event basis ($m=1$), the minimum number of constraint equations clearly implies $k = 2$. With the LSP's now massive we may take our two constraint equations to be $m_{X_2} = m_{Y_2}$ and $m_{X_1} = m_{Y_1}$. Substituting the second expresion into the first reduces the problem to solving one equation for one unknown $m_{X_2} (m_{X_1}) = m_{Y_2} (m_{X_1})$. Expressions for the masses are given by Eq. [\ref{masses2}] with solutions for $c_1$ and $c_2$ given by Eq. [\ref{c2}].

\[
m_{X_2}^2 = m_{X_1}^2 + m_{a_1}^2 + 2 E_{a_1} \sqrt{m_{X_1}^2 + c_1^2 + \vec{p}_{a_1}^2 - (\mathbb{P}^{a_1}_{X_2})^2} - 2(\mathbb{P}^{a_1}_{X_2} + c_1)\mathbb{P}^{a_1}_{X_2} + 2 \vec{p}_{a_1}^2
\]
\beq
m_{Y_2}^2 = m_{X_1}^2 + m_{b_1}^2 + 2 E_{b_1} \sqrt{m_{X_1}^2 + c_2^2 + \vec{p}_{b_1}^2 - (\mathbb{P}^{b_1}_{Y_2})^2} - 2(\mathbb{P}^{b_1}_{Y_2} + c_2)\mathbb{P}^{b_1}_{Y_2} + 2\vec{p}_{b_1}^2
\label{masses2}
\eq

Here we assume that particles $a_1$ and $b_1$ are massive, as per our next example where we study a more general GMSB scenario with massive SM particles and a massive gravitino. The process under consideration will be one in which two partons collide to pair produce two right-handed squarks. The squarks then decay to Higgsino-like neutralinos, labeled by $X_2$ and $Y_2$, emitting jets in the process. The neutralinos then decay to a $Z$ bosons, corresponding to particles $a_1$ and $b_1$, as well as a pair of massive gravitinos $X_1$ and $Y_1$ \cite{Meade:2009qv}. We select events in which each $Z$ boson decays promptly to two leptons so that the intersection of the lepton tracks gives the location of the displaced vertex. Because of the extreme precision with which the detectors can track leptons, this should be the scenario in which secondary vertices may be located with the highest degree of precision. The spectrum for our toy model is given by the following table:

\begin{center}
\begin{tabular}{|c|c|c|}
\hline
Particle & Symbol & Mass \\
\hline
Higgsino & $\tilde{H}$ & $196$ GeV \\
Gravitino & $\tilde{G}$ & $50$ GeV \\
\hline
\end{tabular}
\end{center}

Although the constraint equation $m_{X_2}^2 = m_{Y_2}^2$ is highly non-linear and may have multiple solutions, it can be solved relatively easily using numerical techniques. Unforunately, the existence of multiple solutions may necessitate a larger data sample in order to perform a confident mass measurement. Once the equation has been solved, a numerical value for $m_{X_1}$ can be extracted and used to solve for the exact value of $m_{X_2}$. Here we show a table of the solutions from 5 events with correct and erroneous solutions separated by columns:

\begin{center}
\begin{tabular}{|c|c|c|}
\hline
Event & correct (Gravitino, Higgsino) &  wrong (Gravitino, Higgsino)  \\
\hline
1 & (50, 196)& Null \\
2 & (50, 196)& Null \\
3 & (50, 196)& (120, 287) \\
4 & (50, 196)& (24145, 24349) \\
5 & (50, 196)& Null \\
\hline
\end{tabular}
\end{center}

\noindent Here we see that in this case, gravitino and slepton masses are determined precisely. Though some events evidently contain multiple solutions, the unphysical solutions are sufficiently dispersed about the parameter space so as not to cause confusion in the presence of multiple events when a unique common value can easily be determined by eye. %We can see this again in the following scatter plot where correct answers are denoted by a circle and erroneous solutions are denoted with a $\times$

\subsection{Measurable Displaced Tracks}

The situation is more challenging if particles $a_1$ and $b_1$ are stable as in Figure [\ref{casetwo1}]. If this is the case, then displaced vertices will not be measurable and only the trajectories of the displaced tracks may be observed. This will allow us to subtract only $2m$ from Eq. [\ref{massivecount}], reducing the condition for total kinematic recovery to $4m - 2km + k \leq 0$. Solving for $k$ in terms of $m$ gives $k = 4m/(2m-1)$ so as $m \rightarrow \infty$ we see that $k \rightarrow 2$. Thus the minimum number of constraint equations we can demand is $k=3$, which can be solved using two $m=2$ events. The constraints $m_{X_1} = m_{Y_1}$ and $m_{X_2} = m_{Y_2}$ were combined in Eq. [\ref{masses2}], so the one additional constraint we require for $k=3$ is the condition $m_{X_3} = m_{Y_3}$. The equations for these masses are given in Eq. [\ref{masses3}]

\[
m_{X_3}^2 = m_{X_1}^2 + m_{a_1}^2 + 2 (E_{a_1} + E_{a_2}) \sqrt{m_{X_1}^2 + c_1^2 - (\mathbb{P}^{a_1}_{X_2})^2 + \vec{p}_{a_1}^2} \] \[ - 2(c_1 + \mathbb{P}^{a_1}_{X_2}) (\mathbb{P}^{a_1}_{X_2} + \mathbb{P}^{a_2}_{X_2}) + 2 \vec{p}_{a_1}^2 + 2 E_{a_1} E_{a_2}
\]
\[
m_{Y_3}^2 = m_{Y_1}^2 + m_{b_1}^2 + 2 (E_{b_1} + E_{b_2}) \sqrt{m_{Y_1}^2 + c_2^2 - (\mathbb{P}^{b_1}_{Y_2})^2 + \vec{p}_{b_1}^2} \]
\beq - 2(c_2 + \mathbb{P}^{b_1}_{Y_2}) (\mathbb{P}^{b_1}_{Y_2} + \mathbb{P}^{b_2}_{Y_2}) + 2 \vec{p}_{b_1}^2 + 2 E_{b_1} E_{b_2} 
\label{masses3}
\eq

These equations are again calculated assuming massive $a_1$ and $b_1$ as per our example, though the assumption of massless particles $a_2$ and $b_2$ is still taken for simplicity. It should be noted however, that all equations generalize easily to arbitrary massive SM particles. From the above equations we see that it is possible to construct expressions for $m_{X_2}$, $m_{Y_2}$, $m_{X_3}$ and $m_{Y_3}$ in terms of three unknown quantities $z_X$, $z_Y$ and $m_{X_1}$. For each event we have two constraint equations $m_{X_1} = m_{Y_1}$ and $m_{X_2} = m_{Y_2}$ with which to solve them. First notice that since the LSP 4-momenta can be calculated in terms of $z_X$, $z_Y$ and $m_{X_1}$, we can explicitly express $m_{X_2}$ and $m_{X_3}$ in terms of these variables. This means that we can apply a change of variables and parameterize the three unknown quantities instead as $m_{X_1}$, $m_{X_2}$ and $m_{X_3}$. The fact that there are two constraint equations means that the solution for each event is a curve in parameter space, which in this case is just $\mathbb{R}^3$ with axes labeled ($m_{X_1}, m_{X_2}, m_{X_3}$). Since the trajectories of the displaced tracks are unique for each event, curves generated by different events will be unique but will always traverse the correct answer. Thus in principle, the correct value for the masses will exist at the intersection of the curves, which is clearly equivalent to the condition of matching particle masses from different events.

Put another way, every hypothesis for the value of $m_{X_1}$ is equivalent to a hypothesis for the values of $z_X$ and $z_Y$. It is thus also equivalent to a hypothesis for the values of $m_{X_2}$ and $m_{X_3}$. By considering a range of hypotheses for $m_{X_1}$ over a few events, the correct values of $m_{X_2}$ and $m_{X_3}$ will be the unique intersection of all hypotheses. A demonstration of this scenario has been performed with the following mass spectrum (the Bino and Slepton masses are the same as before but the Gravitino mass is now set to 50 GeV):

\begin{center}
\begin{tabular}{|c|c|c|}
\hline
Particle & Symbol & Mass \\
\hline
Bino & $\tilde{B}$ & $199$ GeV \\
Right-handed Slepton & $\tilde{l}_R$ & $107$ GeV \\
Gravitino & $\tilde{G}$ & $50$ GeV \\
\hline
\end{tabular}
\end{center}

\begin{figure}[!ht]
\begin{center}
\includegraphics[width=100mm]{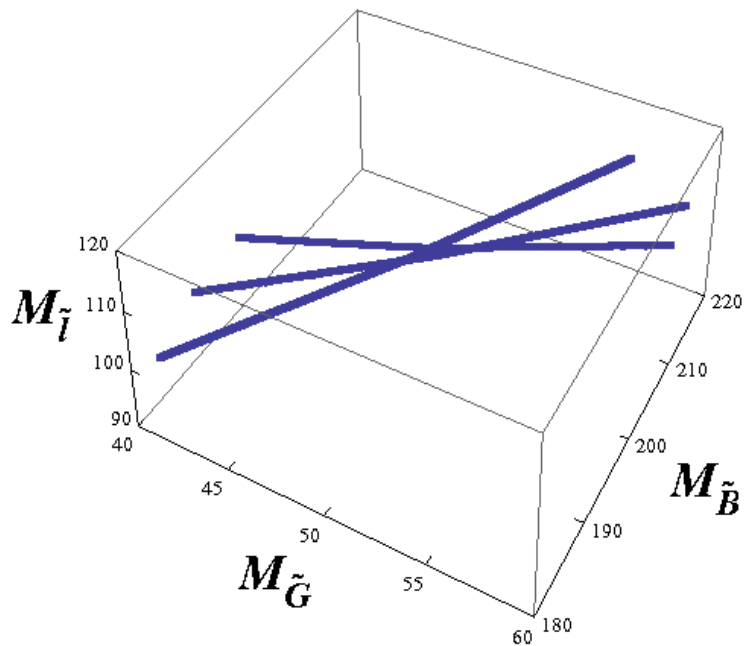}
\caption{With three unknowns ($m_{\tilde{B}}$, $m_{\tilde{l}}$, $m_{\tilde{G}}$) and two equations, the solutions are curves in three dimensional Euclidean space. The intersection of solutions should occur at the correct value of the masses as can be seen in this plot using 3 events as an example.}
\label{interlines}
\end{center}
\end{figure}

As explained, the parameter space for this scenario is $\mathbb{R}^3$ with axes labeled $(m_{\tilde{B}},m_{\tilde{l}},m_{\tilde{G}})$. Analyzing three events, we scan values of the gravitino mass from $0$ to $100$ GeV. As expected, this scan produces a curve in parameter space for each event with the correct answer lying at the intersection of the curves as shown in Figure [\ref{interlines}]. For reasons given earlier, in practice we do not expect an exact intersection, but rather a localized region in parameter space where the density of such lines achieves a maximum. The optimal method of mass extraction should then involve searching for the slice in the $m_{\tilde{G}}$ plane where the density of solutions for $m_{\tilde{B}}$ and $m_{\tilde{l}}$ achieves a maximum. To this end we compute a probability sum on each slice of equal $m_{\tilde{G}}$ using the Gaussian distribution in Eq. [\ref{gaussfit2}] as our probability distribution function with $\sigma = 10$ GeV.

\beq
F[m_{\tilde{B},0},m_{\tilde{l},0}] =\frac{1}{(2 \pi \sigma)^2} \displaystyle\sum_i \exp \bigg( -\frac{(m_{\tilde{B},i}-m_{\tilde{B},0})^2+(m_{\tilde{l},i}-m_{\tilde{l},0})^2}{2\sigma^2} \bigg)
\label{gaussfit2}
\eq

\begin{figure}[!ht]
\begin{center}
\includegraphics[width=100mm]{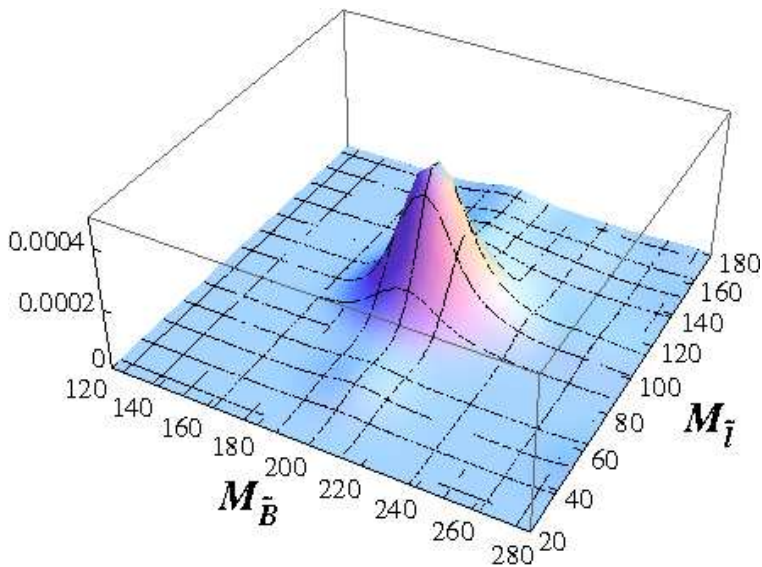}
\caption{The probability sum on the $m_{\tilde{G}} = 50$ GeV slice. We see that it peaks at the correct value of $m_{\tilde{B}}$ and $m_{\tilde{l}}$}
\label{likeright}
\end{center}
\end{figure}

\begin{figure}[!ht]
\begin{center}
\includegraphics[width=100mm]{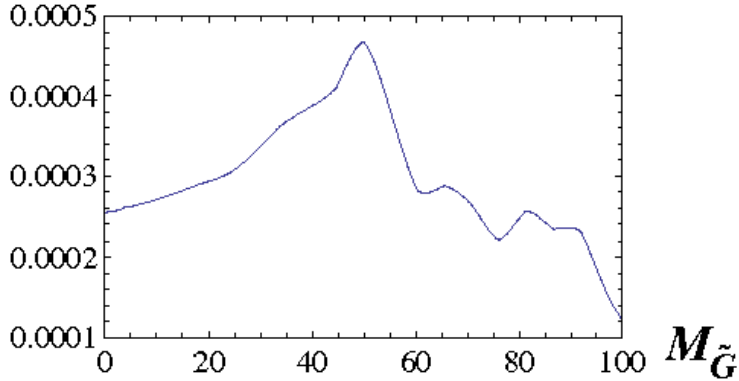}
\caption{A plot of the absolute height of the probability sum as a function of $m_{\tilde{G}}$. We see that it peaks at the correct value.}
\label{likefin}
\end{center}
\end{figure}

Slices of equal $m_{\tilde{G}}$ give a plane parameterized by $m_{\tilde{B}}$ and $m_{\tilde{l}}$. The function in Eq. [\ref{gaussfit2}] is defined at each point on this plane ($m_{\tilde{B},0}$, $m_{\tilde{l},0}$) and the sum over $i$ takes a contribution from each data point ($m_{\tilde{B},i}$, $m_{\tilde{l},i}$), which is just given by the intersection of each line with the equal $m_{\tilde{G}}$ slice. Therefore, Eq. [\ref{gaussfit2}] should be maximixed at the point in the plane with the highest density of solutions. Furthermore, the maximum height in each plane should achieve its largest absolute magnitude on the slice corresponding to the correct value of $m_{\tilde{G}}$, since it is on this plane that the highest density of solutions resides. Using a sample of 25 events, the probability sum on the correct $m_{\tilde{G}}$ slice is depicted in Figure [\ref{likeright}] and we can observe a clear maximum at the correct solution for $m_{\tilde{B}}$ and $m_{\tilde{l}}$. The maximum height for each $m_{\tilde{G}}$ is then plotted as a function of the $m_{\tilde{G}}$ in Figure [\ref{likefin}]. As expected, the largest absolute magnitude for the probability sum is acheived at the correct value of $m_{\tilde{G}} = 50$ GeV.

\subsection{Can We Do Better?}

As we saw in the previous section, as the number of events $m \rightarrow \infty$, the minimum number of constraint equations needed $k \rightarrow 2$. Thus naively it would seem impossible to solve for all particle masses using the condition $k=2$, with a topology given in Figure [\ref{casethree1}]. Previously we saw that it was still possible to measure the masses in situations where the unknown quantities outnumbered the constraints, using a very small number of events, by employing the trick of section $4.3$. It is thus sensible to ask the question of whether or not it would be possible to perform an analagous measurement on cascade decays with a massive LSP. The spectrum used for this example was as follows:

\begin{center}
\begin{tabular}{|c|c|c|}
\hline
Particle & Symbol & Mass \\
\hline
Right-handed Slepton & $\tilde{l}_R$ & $107$ GeV \\
Gravitino & $\tilde{G}$ & $50$ GeV \\
\hline
\end{tabular}
\end{center}

\begin{figure}[!ht]
\begin{center}
\includegraphics[width=100mm]{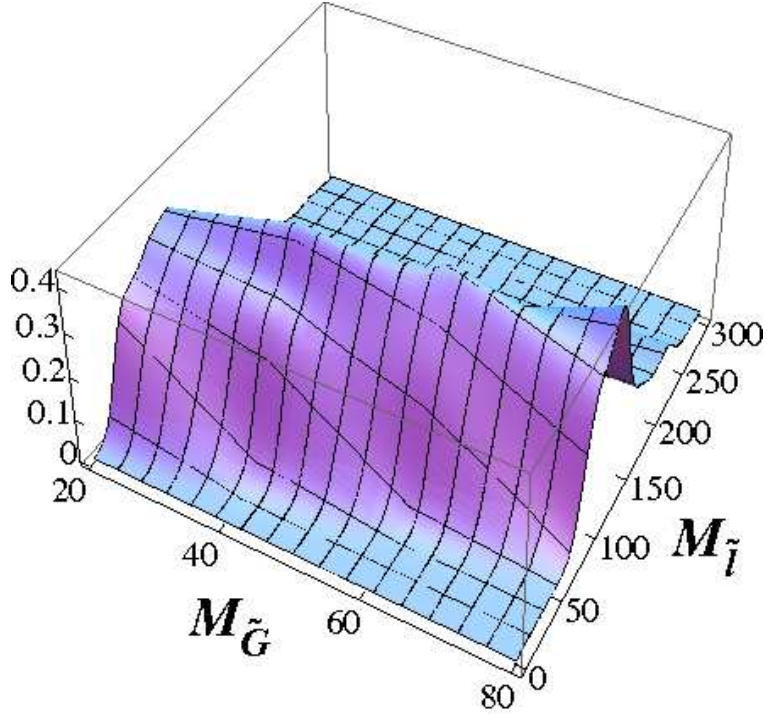}
\caption{The probability double sum for the massive LSP case where unknowns outnumber constraints. The ridge-like structure suggests that no unique solutions exists for the correct value of $m_{\tilde{G}}$ using this technique}
\label{ridge}
\end{center}
\end{figure}

As usual, with a massive LSP we have 4 unknowns which we may take to be $z_X$, $z_Y$, $m_{X_1}$ and $m_{Y_1}$. Since we are assuming $k=2$, the available kinematic equations can only remove two unknowns. In analogy with the massless LSP scenario, we choose to eliminate the two LSP masses and can derive an expression for the mass of the NLSP $m_{X_2} \rightarrow m_{X_2} (z_X, z_Y)$.  We now scan all possible values for $m_{\tilde{G}}$ and play the same trick that was used in the previous section, but in one higher dimension. The result is depicted in Figure [\ref{ridge}]. Unfortunately, this ``probability-double-sum'' produces a ridge-like structure rather than a peak at the correct solution. This result implies that these techniques cannot be used to extract a unique solution for $m_{\tilde{G}}$ when it is non-zero, and can only be used to provide a relation between two mass parameters.

\section{Conclusions}

In this paper we studied scenarios in which heavy new particle states were pair produced and cascaded down to some non-interacting stable particle states generating visible SM particles along the way.  Here we assumed the decay length of the last decay was measurable, which resulted in a signature of displaced vertices or tracks. We finally assumed that the LSP's were the only particles that contributed to the transverse missing momentum. Given these assumptions, we described a number of novel techniques for extracting the spectrum of the intermediary particles in the cascade decay that were effective even in the low statistics limit. They would therefore be useful for very early discovery level searches at the LHC.

It should obvious by now that although this procedure is completely model independent, it was inspired by the phenomenology of supersymmetric models. For any supersymmetric theory on which these methods may be applied, the following conditions must hold:

\begin{enumerate}

\item R-parity must be (approximately) preserved guaranteeing the stability of the LSP
\item Decays to the LSP must occur before the cascade reaches the detector
\item The 4-momentum of the NLSP must be traceable back to the primary vertex
\item The LSP's must be the sole source of missing transverse momentum
\item The decay length of the NLSP must be finite and measurable

\end{enumerate}

The first four assumptions are very generic for SUSY models though the fifth assumption is rather specific. Despite this fact it can be generically realized in many models, providing us with additional handles on the kinematics of these events. In our paper, we focus on the scenario in which the final step of decay happens at a reasonable finite distance but before the NLSP hits the detector. In scenarios with gauge-mediated SUSY breaking, the decay length of the NLSP is directly related to NLSP mass and the SUSY breaking scale via the relation $(c \tau)_{NLSP} \sim (\sqrt{F})^4 / m_{X_2}^5$. Since all of the techniques presented in this paper also provide a direct measurement of the decay length of the NLSP, if SUSY is realized in nature they could also be used to extract a very early measurement of the SUSY-breaking scale.

Recently there has been a lot of talk about optimizing search strategies for very early discovery level analyses at the LHC. A central theme in these discussions has been the idea of ``searching under the lamp post''. The principle behind this theme is that at the very early stages of a new physics search, especially when data is sparse and statistics are low, it may be a better strategy to search for that which is easiest to see rather than that which you think is most likely to be true. If new physics manifests itself through the presence of missing energy and dual displaced tracks, with $\mathcal{O} (few)$ events these techniques provide the possibility of

\begin{enumerate}

\item Providing convincing evidence for the existence of dual cascade decay topologies
\item Measuring the masses of all new particle states participating in the cascade decay
\item Constructing accurate distributions illuminating the spin-structure of the particles
\item Calculating the SUSY breaking scale if nature is supersymmetric

\end{enumerate}

\noindent Clearly the methods described in this paper allow for a very large return from a very small investment. In particular, they allow one to extract an enormous amount of information from signatures that would otherwise be left to very late post-discovery analyses to elucidate completely. As such, they present an extremely bright lamp post under which to search in the coming months.

\section{Appendix A}

Here we give some details about how to prepare the MC information and how we derive the weighting factor.

We first took the lhe file from a fixed SUSY spectrum, where the 4-momenta of gravitino, slepton and lepton are accessible. The information of the beginning and ending points of displaced tracks are missing.  We take the proper decay length of slepton to be half of detector radius and impose the location of 2nd vertex according to the exponential decay distribution.  With all those information, we can calculate the beginning and ending points of displaced tracks for each event. Since displaced tracks are assumed to be measurable in experiments, but 4-momenta of gravitino and slepton are not, we would use the track information and forget the momenta information for the later analysis.  Also, the transverse missing energy is assumed to be only coming from two gravitinos, thus it can be calculated when we prepare the MC information.  One needs to be very careful on what information is accessible and what is not. We summarize the accessible information as following: the location of primary vertex, the beginning and ending points of the two displaced tracks for each event, the 4-momenta of each displaced track, and transverse missing energy. Except for those, all other information will be treated as inaccessible.

After we finish the preparation of MC information, we proceed the scanning in possible location of 2nd vertex along one of the displaced tracks.  The scanning we do is taking a constant step along the track.  The calculation of weighting factor for each scanning point is a little tricky, because part of the information of exponential decay has already been included in the distribution of displaced tracks.  Thus the uniform step of scanning along each tracks in many events is not giving us a uniform distribution of points in space. To gain the correct weighting factor, one has to first include a weighting factor of each displaced track. This factor is described more precisely in the following Appendix, then an exponential decay factor can be applied. Thus the weighting factor for each point takes the form in Eq. [\ref{weightfunc}]

\section{Appendix B}

\begin{figure}[!ht]
\begin{center}
\includegraphics[width=100mm]{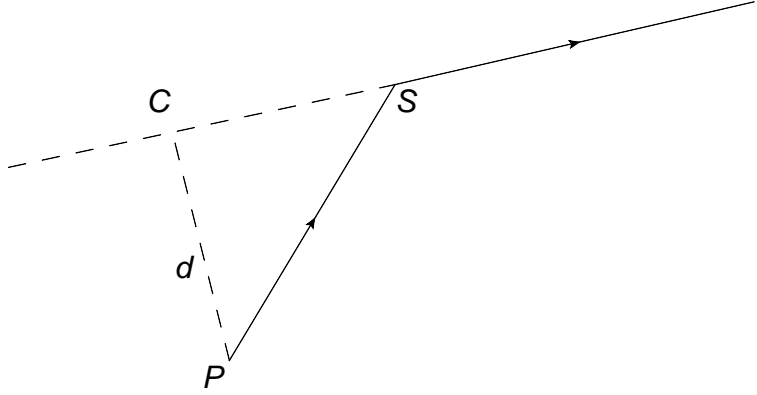}
\caption{The kinematics of a displaced track. Here $d$ is distance of closest approach between the primary vertex and the displaced track, $C$ is the point of closest approach, $l$ is the distance between the primary vertex and the hypothetical point of decay on the displaced track and $l_0$ is the characteristic decay length of the NLSP.}
\label{secvert}
\end{center}
\end{figure}

Here we give a detailed discussion on how to extract a rough estimate of the decay length from the observation of a few displaced tracks. This can be done by looking at the distribution of perpendicular distances from the primary vertex to the point of closest approach for all displaced tracks $d_{\perp}$. Suppose $X_2$ travels a distance $l$ before decaying, as in Figure [\ref{secvert}]. Let $\theta$ be the angle between PS and CS. Then we have

$$d_{\perp}=l \sin \theta $$

\noindent Let $l_0$ be the characteristic decay length of particle $X_2$.Since the measured decay distance for an event $l$ must  an exponentially decaying distribution, the properly normalized probability distribution is:

$$\frac{\partial P(l,\theta,\phi)}{\partial l}|_{\theta,\phi}=\frac{1}{l_0}e^{-l/l_0}F(\theta,\phi)$$

\noindent Also, we know that the decay of particle $X_2$ is isotropic in its rest frame. Since the mass of $X_2$ is $O(100 GeV)$, the boost from lab frame to the rest frame of $X_2$ is not large. We can thus approximate the anglular distribution in lab frame to be isotropic:

$$\frac{\partial^2 P(l,\theta,\phi)}{\partial (\cos \theta) \partial\phi}|_{l}=\frac{1}{4\pi}G(l)$$

\noindent Since $l$, $\theta$ and $\phi$ are independent variables, we :

$$\frac{\partial^3 P(l,\theta,\phi)}{\partial l \partial (\cos \theta) \partial\phi}=\frac{1}{4\pi l_0}e^{-l/l_0}$$

\noindent From the relation between $d_{\perp}$ and $(l,\theta)$, we
have

$$\frac{\partial d_\perp}{\partial(\cos \theta)}|_l=-l\sqrt{\frac{l^2}{d_\perp^2}-1}$$

\noindent Finally, we get

$$g(d_{\perp})=\frac{d P}{d(d_\perp)}=\int_{d_{\perp}}^{\infty}\frac{dl}{2l_0}\frac{d_{\perp}e^{-l/l_0}}{l\sqrt{l^2-d_{\perp}^2}}$$

\noindent Though this integral is not easy to solve, one can cut the integral at very large values and get the distribution numerically. Thus, with just a couple of displaced tracks from a few events, one can extract the rough value of the decay length of $X_2$.

\section{Acknowledgements}
We would like to thank Simon Knapen and Michael Peskin for very helpful discussions. Also special thanks to Scott Thomas for the continued support and the many helpful suggestions. The research of MP and YZ is supported in part by the NHETC in the Department of Physics and Astronomy at Rutgers University, and in part by DOE grant DE-FG02-96ER40949.

% --------------------------------------------------------------------

\end{document}